\documentstyle[12pt]{article}
\newlength{\extraspace}
\newlength{\extraspaces}
\newcommand{\be}{\begin{equation}
\addtolength{\abovedisplayskip}{\extraspaces}
\addtolength{\belowdisplayskip}{\extraspaces}
\addtolength{\abovedisplayshortskip}{\extraspace}
\addtolength{\belowdisplayshortskip}{\extraspace}}
\newcommand{\ee}{\end{equation}}

\newcommand{\ba}{\begin{eqnarray}
\addtolength{\abovedisplayskip}{\extraspaces}
\addtolength{\belowdisplayskip}{\extraspaces}
\addtolength{\abovedisplayshortskip}{\extraspace}
\addtolength{\belowdisplayshortskip}{\extraspace}}
\newcommand{\ea}{\end{eqnarray}}

\newcommand{\nonu}{\nonumber \\[.5mm]}
\newcommand{\A}{&\!\!\!}

\newcommand{\newsection}[1]{
\vspace{7mm} \pagebreak[3] \addtocounter{section}{1}
\setcounter{subsection}{0} \setcounter{footnote}{0}
\begin{center}
{\large {\bf \thesection. #1}}
\end{center}
\nopagebreak
\medskip
\nopagebreak \hspace{3mm}}

\begin{document}
\begin{center}
{{\bf Energy of spherically symmetric spacetimes on regularizing
teleparallelism}}\footnote{\hspace*{-0.7cm} PACS numbers: 04.20.Cv, 04.20.Fy, 04.50.-h\\
 Keywords: gravitation, teleparallel gravity,
energy-momentum, Weitzenb$\ddot{o}$ck connection, regularization
teleparallelism}
\end{center}
\centerline{Gamal G.L. Nashed\footnote{\hspace*{-0.7cm}Mathematics
Department, Faculty of Science, Ain Shams University, Cairo,
Egypt}}
\bigskip

\centerline{\it Centre for Theoretical Physics, The British
University in Egypt, El-Sherouk City,} \centerline{{\it Misr -
Ismalia Desert Road, Postal No. 11837, P.O. Box 43, Egypt.}}

\bigskip
\centerline{ e-mail:nashed@bue.edu.eg}

\hspace{2cm} \hspace{2cm}
\\
\\
\\
\\
\\
\\
\\
\\
\\

We calculate the total energy of an exact spherically symmetric
solutions, i.e., Schwarzschild and Reissner Nordstr$\ddot{o}$m,
using the gravitational energy-momentum 3-form within the tetrad
formulation of general relativity. We explain how  the effect of
the inertial makes the total energy unphysical! Therefore, we use
the covariant teleparallel approach which makes the energy always
physical one. We also show that the inertial has no effect on the
calculation of  momentum.

\newsection{Introduction}

The best well known gravitational theory is the Einstein one. This
theory up to this day is consistence with observational data. The
geometry in which Einstein general relativity (GR) based on is the
Riemannian geometry with  unique metric and unique connection.
However, due to the geometric structure and the equivalence
principle  of the gravitational theory the problem of  energy is
not completely solved until now. Using the Lagrange-Noether
approach, one can derives  the conserved currents that are arise
from the invariance of the classical action under transformations
of fields. However, in Riemannian geometry one can not find
symmetries that can be used to generate the conserved
energy-momentum currents.  Only one can speak about the energy of
asymptotically  flat spacetime. Earlier analyses of this problem
can be found in  details  in (\cite{Ta}$\sim$\cite{Nj} and
references therein) for example.

As is well known, gravitational interaction can be described
either by curvature or torsion \cite{OP1}.  According to GR,
curvature is used to geometrize spacetime, and in this way a
successful description  to the gravitational interaction is
carried out. On the other hand teleparallelism, attributes
gravitation to torsion. In this case torsion accounts for
gravitation not by geometrizing the interaction, but by acting as
a force. This means that, in the teleparallel equivalent of
general relativity, there are no geodesics, but force equations
quite analogous to the Lorentz force equation of electrodynamics
\cite{AP}. Therefore, Gravitational interaction, can be described
either in terms of curvature, as is usually done in GR, or in
terms of torsion, in which case we have the teleparallel gravity.

Teleparallel theories are interesting for several reasons: first
of all, GR can be viewed as a particular theory of teleparallelism
and, thus, teleparallelism could be considered at the very least
as a different point of view that can lead to the same results
\cite{TM}.  Second, in this framework, one can define an
energy-momentum tensor for the gravitational field that is a true
tensor under all general coordinate transformations. This is the
reason why teleparallelism was reconsidered by M\o ller  when he
was studying the problem of defining an energy-momentum tensor for
the gravitational field \cite{M2}. The idea was taken over by
Pellegrini and Pleba\'{n}ski that constructed the general
Lagrangian for these theories \cite{PP}. The third reason why
these theories are interesting is that they can be seen as gauge
theories of the translation group (not the full Poincar$\acute{e}$
group) and, thus, they give an alternative interpretation of GR
\cite{Hw,HS}.

An important difference between Einstein GR theory and
teleparallel theories  is that it is possible to distinguish
gravitation and inertia \cite{AGP}. Since inertia is in the realm
of the pseudotensor behavior of the usual expressions for the
gravitational energy-momentum density, it turns out possible in
teleparallel gravity to write down a tensorial expression for such
density \cite{ALP}. With the purpose of getting a deeper insight
into the covariant teleparallel formalism, as well as to test how
it works Lucas  et al. \cite{LOP} reanalyze the computation of the
total energy of two  examples. Recently Obukhov   et al.
\cite{OPR} computed the energy and momentum transported by exact
plane gravitational-wave solutions of Einstein equations using the
teleparallel equivalent formulation of Einstein's theory. It is
our aim to extend the calculations done by  Lucas et al.
\cite{LOP} using Schwarzschild and Reissner Nordstr$\ddot{o}$m
solutions  with local Lorentz transformations contain two
constants $c_1$ and $c_2$. Also we  show how is inertia related to
Weitzenb$\ddot{o}$ck connection\footnote{We will use the same
notation given in Ref. \cite{LOP}}.

In \S 2, we use the language of exterior forms to give  an outline
of the teleparallel approach. A brief review of the covariant
formalism for the gravitational energy-momentum  which is
described by the pair $( \vartheta^{\alpha},
{\Gamma_\alpha}^\beta)$ is also given in  \S 2.  In \S 3, we show
by calculations that due to an inconvenient choice of a reference
system, the traditional computation of the total energy of
Schwarzschild and  Reissner Nordstr$\ddot{o}$m  solutions are
unphysics! Using the  covariant formalism, we show that the
Weitzenb$\ddot{o}$ck connection acts as a regularizing tool that
separates the inertial contribution and provides the physically
meaningful result. Final section is devoted for main results and
discussion.

We use the Latin indices ${\it i, j, \cdots }$ for local holonomic
spacetime coordinates and the Greek indices $\alpha$, $\beta$,
$\cdots$ label (co)frame components. Particular frame components
are denoted by hats, $\hat{0}$,$\hat{1}$, etc. As usual, the
exterior product is denoted by $\wedge$, while the interior
product of a vector $\xi$ and a p-form $\Psi$ is denoted by $\xi
\rfloor \Psi$. The vector basis dual to the frame 1-forms
$\vartheta^{\alpha}$ is denoted by $e_\alpha$ and they satisfy
$e_\alpha \rfloor \vartheta^{\alpha}={\delta_\alpha}^\beta$. Using
local coordinates $x^i$, we have $\vartheta^{\alpha}=h^\alpha_i
dx^i$ and $e_\alpha=h^i_\alpha \partial_i$ where $h^\alpha_i$ and
$h^i_\alpha $ are the covariant and contravariant components of
the tetrad field. We define the volume 4-form by $\eta \stackrel
{\rm def.}{=} \vartheta^{\hat{0}}\wedge \vartheta^{\hat{1}}\wedge
\vartheta^{\hat{2}}\wedge\vartheta^{\hat{3}}.$  Furthermore, with
the help of the interior product we define \[\eta_\alpha \stackrel
{\rm def.}{=} e_\alpha \rfloor \eta = \ \frac{1}{3!} \
\epsilon_{\alpha \beta \gamma \delta} \ \vartheta^\beta \wedge
\vartheta^\gamma \wedge \vartheta^\delta,\]  where
$\epsilon_{\alpha \beta \gamma \delta}$ completely antisymmetric
with $\epsilon_{0123}=1$. \[\eta_{\alpha \beta} \stackrel {\rm
def.}{=} e_\beta \rfloor \eta_\alpha =
\frac{1}{2!}\epsilon_{\alpha \beta \gamma \delta} \
\vartheta^\gamma \wedge \vartheta^\delta,\]
\[\eta_{\alpha \beta \gamma} \stackrel {\rm def.}{=} e_\gamma
\rfloor \eta_{\alpha \beta}= \frac{1}{1!} \epsilon_{\alpha \beta
\gamma \delta} \  \vartheta^\delta,\]  which are bases for 3-, 2-
and 1-forms respectively. Finally, \[\eta_{\alpha \beta \mu \nu}
\stackrel {\rm def.}{=} e_\nu \rfloor \eta_{\alpha \beta \mu}=
e_\nu \rfloor e_\mu \rfloor e_\beta \rfloor e_\alpha \rfloor
\eta,\] is the Levi-Civita tensor density. The $\eta$-forms
satisfy the useful identities: \ba \vartheta^\beta \wedge
\eta_\alpha \A \stackrel {\rm def.}{=}  \A \delta^\beta_\alpha
\eta, \nonu
\vartheta^\beta \wedge \eta_{\mu \nu} \A \stackrel {\rm def.}{=}
\A \delta^\beta_\nu \eta_\mu-\delta^\beta_\mu \eta_\nu, \nonu
\vartheta^\beta \wedge \eta_{\alpha \mu \nu} \A \stackrel {\rm
def.}{=}  \A \delta^\beta_\alpha \eta_{\mu \nu}+\delta^\beta_\mu
\eta_{\nu \alpha}+\delta^\beta_\nu \eta_{ \alpha \mu}, \nonu
\vartheta^\beta \wedge \eta_{\alpha \gamma \mu \nu} \A \stackrel
{\rm def.}{=}  \A \delta^\beta_\nu \eta_{\alpha \gamma
\mu}-\delta^\beta_\mu \eta_{\alpha \gamma \nu
}+\delta^\beta_\gamma \eta_{ \alpha \mu \nu}-\delta^\beta_\alpha
\eta_{ \gamma \mu \nu}. \ea

The line element $ds^2 \stackrel {\rm def.}{=} g_{\alpha \beta}
\vartheta^\alpha \bigotimes \vartheta^\alpha$ is defined by the
spacetime metric $g_{\alpha \beta}$.

\newsection{Berif review of teleparallel gravity and  energy-momentum conservation}
Teleparallel geometry can be viewed as a gauge theory of
translation \cite{Hw,HS,Cy}$\sim$ \cite{Tr}. The coframe
$\vartheta^\alpha$  can be viewed as a one-form that plays the
role of the gauge translational potential of the gravitational
field. Einstein's general relativity theory can be reformulated as
teleparallel equivalent to GR theory. Geometrically, one can view
the teleparallel gravity as a special (degenerate) case
\cite{OP1,OP2,ORP} of the metric-affine gravity in which the
coframe  $\vartheta^\alpha$ and the local Lorentz connection
${\Gamma_\alpha}^\beta$ are subject to the distant parallelism
constraint ${R_\alpha}^\beta=0$ \cite{OP1}. The torsion 2-form \be
T^\alpha\stackrel {\rm def.}{=} d\vartheta^\alpha+
{\Gamma_\beta}^\alpha\wedge \vartheta^\beta,\ee arises as the
gravitational gauge field strength and ${\Gamma_\alpha}^\beta$
being the Weitzenb$\ddot{o}$ck connection. The torsion $T^\alpha$
can be decomposed into three irreducible pieces: the tensor part,
the trace, and the axial trace, given respectively by, \cite{LOP},
for example \ba {^ {\tiny{( 1)}}T^\alpha}  \A \stackrel {\rm
def.}{=} \A T^\alpha-{^ {\tiny{( 2)}}T^\alpha}-{^  {\tiny{(
3)}}T^\alpha}, \qquad with \nonu
{^  {\tiny{( 2)}}T^\alpha}  \A \stackrel {\rm def.}{=} \A
\frac{1}{3} \vartheta^\alpha\wedge \left(e_\beta \rfloor
T^\beta\right),\nonu
{^  {\tiny{( 3)}}T^\alpha}  \A \stackrel {\rm def.}{=} \A
\frac{1}{3} e^\alpha\rfloor \left(\vartheta^\beta \wedge
T_\beta\right).\ea The Lagrangian of the teleparallel equivalent
gravity model reads \be V = -\frac{1}{2\kappa}T^\alpha  \wedge
^\ast \left({^  {\tiny{( 1)}}T_\alpha}-2{^  {\tiny{( 2)}}T_\alpha}
-\frac{1}{2}{^  {\tiny{( 3)}}T_\alpha} \right). \ee $\kappa=8\pi
G/c^3$, where $G$ is the Newtonian constant and $c$ is the speed
of light, $\ast$ denotes the Hodge duality in the metric
$g_{\alpha \beta}$ which is assumed to be flat Minkowski metric
$g_{\alpha \beta}=o_{\alpha \beta}=diag(+1,-1,-1,-1)$, that is
used to raise and lower local frame (Greek) indices.

The teleparallel field equations are obtained from the variation
of the total action with respect to the coframe \be
DH_\alpha-E_\alpha=\Sigma_\alpha,\ee where   $DH_\alpha=
dH_\alpha-\Gamma_\alpha^\beta \wedge H_\beta$ denotes   the
covariant exterior derivative and $\Sigma_\alpha$ is the canonical
energy-momentum current 3-form of matter \be \Sigma_\alpha
\stackrel {\rm def.}{=} \frac{\delta L_{mattter}}{\delta
\vartheta^\alpha} \ee as the source. In accordance with the
general Lagrange-Noether scheme \cite{Gf, HMM}  one derives from
(4) the translational momentum 2-form and the canonical
energy-momentum 3-form: \be H_{\alpha} \stackrel {\rm def.}{=}
-\frac{\partial V}{\partial T^\alpha}=\frac{1}{\kappa} \ast
\left({^  {\tiny{( 1)}}T_\alpha}-2{^  {\tiny{( 2)}}T_\alpha}
-\frac{1}{2}{^  {\tiny{( 3)}}T_\alpha} \right), \ee \be E_\alpha
\stackrel {\rm def.}{=} \frac{\partial V}{\partial
\vartheta^\alpha}=e_\alpha \rfloor V+\left(e_\alpha \rfloor
T^\beta \right) \wedge H_\beta. \ee Due to geometric identities
\cite{Oyn}, the gauge momentum (4) can be recast as \be
V=-\frac{1}{2}T^\alpha\wedge H_\alpha.\ee The model resulting from
the Lagrangian (4) is degenerate from the metric-affine viewpoint,
because the variational derivatives of the action with the respect
to the metric and connection are trivial. This means that the
field equations
are satisfied for any  ${\Gamma^\alpha}_\beta$. The presence of the
connection field plays an important regularizing role as shown in \cite{LOP}.
 The latter is twofold:\\
\underline{First}: The teleparallel gravity becomes explicitly
covariant under the local Lorentz transformations of the coframe.
In particular, the Lagrangian (4) is invariant under the change of
variables \be   \vartheta'^\alpha={L^\alpha}_\beta
\vartheta^\beta, \qquad
{\Gamma'_\alpha}^\beta={\left(L^{-1}\right)^\mu}_\alpha
{\Gamma_\mu}^\nu {L^\beta}_\nu+{L^\beta}_\gamma
d{\left(L^{-1}\right)^\gamma}_\alpha,\ee where
${L^\alpha}_\beta(x) \in SO(1,3)$. In the pure tetrad gravity
which can be recovered when ${\Gamma_\alpha}^\beta=0$, the
Lagrangian is only quasi-invariant—it changes by a total
divergence.

The connection  ${\Gamma_\alpha}^\beta$ can be decomposed into
Riemannian and post-Riemannian parts as \be {\Gamma_\alpha}^\beta
\stackrel {\rm def.}{=} {\tilde{\Gamma}_\alpha}^\beta
-K_{\alpha}^\beta,\ee with ${\tilde{\Gamma}_\alpha}^\beta $  is
the purely Riemannian connection and $K^{\mu \nu}$ is the
contorsion 1-form which is related to the torsion through the
relation

\be T^\alpha  \stackrel {\rm def.}{=}  {K^\alpha}_\beta \wedge
\vartheta^\beta.\ee The translational momentum (7) can be
rewritten as \cite{Oyn} \be H_\alpha=\frac{1}{2\kappa} K^{\mu
\nu}\wedge \eta_{\alpha \mu \nu}.\ee \underline{Second}: The more
important property of the teleparallel framework is that the
Weitzenb$\ddot{o}$ck connection actually represents inertial
effects that arise due to the choice of the reference system
\cite{ALP}. The inertial contributions in many cases yield
unphysical results for the total energy of the system, producing
either trivial or divergent answers. The teleparallel connection
acts as a regularizing tool which helps to subtract the inertial
effects without distorting the true gravitational contribution
\cite{LOP}.

In the Maxwell-type the field equation (5) can be rewritten in the
form \be DH_\alpha=E_\alpha+\Sigma_\alpha.\ee The Maxwell 2-form
$F=dA$ represents the gauge field strength of the electromagnetic
potential 1-form A. Using  the Lagrangian $V(F)$, the 2-form of
the electromagnetic excitations is defined by $H=-\frac{\partial
V}{\partial F}$. The field equations has the form $dH=J$ where $J$
is the 3-form of the electric current density of matter. In view
of the nilpotency of the exterior differential, $dd=0$, the
Maxwell equation yields the conservation law of the electric
current, $dJ= 0$.

Similarly to electrodynamics, gravity is a self-interacting field,
and the gauge field potential 1-form $\vartheta^\alpha$ carries an
``$\,$ internal" index $\alpha$. The gauge field strength 2-form
$T^\alpha= D\vartheta^\alpha$ is now defined by the covariant
derivative of the potential (compare with $F = dA$). The
gravitational field excitation 2-form $H_\alpha$ introduced by
(7), in a direct analogy to the Maxwell theory ($
H=-\frac{\partial V}{\partial F}$). Finally, we observe that as
compared to the Maxwell field equation $dH = J$, the gravitational
field equation (14) contains now the covariant derivative $D$, and
in addition, the right-hand side is represented by a modified
current 3-form, $E_\alpha+\Sigma_\alpha$. The last term is the
energy-momentum of matter, and we naturally conclude that the
3-form $E_\alpha$ describes the energy-momentum current of the
gravitational field. Its presence in the right-hand side of the
field equation (14) reflects the self-interacting nature of the
gravitational field, and such contribution is absent in the linear
electromagnetic theory.

Comparison with electrodynamic can be completed by deriving the
corresponding conservation law. Indeed, since $DD \equiv 0$ for
the teleparallel connection, (14) tells us that the sum of the
energy-momentum currents of gravity and matter,
$E_\alpha+\Sigma_\alpha$, is covariantly conserved \cite{ALP}, \be
D\left(E_\alpha+\Sigma_\alpha\right)=0.\ee This law is consistent
with the covariant transformation properties of the currents
$E_\alpha$ and $\Sigma_\alpha$.

One can rewrite the conservation of energy-momentum in terms of
the ordinary derivatives. Using the explicit expression
$DH_\alpha=dH_\alpha-{\Gamma_\alpha}_\beta \wedge H_\beta,$  the
field equation (5) and (14) can be recast in an alternative form
\be d\left(\varepsilon_\alpha+\Sigma_\alpha\right)=0.\ee

The 3-form $E_\alpha$ describes the gravitational energy-momentum
in a covariant way, whereas the 3-form $\varepsilon_\alpha$ is a
noncovariant object. In terms of components, it gives rise to the
energy-momentum pseudotensor. It is worthwhile to note that
$H_\alpha$ plays the role of energy-momentum superpotential both
for the covariant energy-momentum current
$\left(E_\alpha+\Sigma_\alpha\right)$ and for the total (including
inertia) non-covariant current
$\left(\varepsilon_\alpha+\Sigma_\alpha\right)$.

The $\eta$-forms defined defined in Eq. (1) serve as the basis of
the spaces of forms of different rank, and when we expand the
above objects with respect to the  $\eta$-forms, the usual tensor
formulation is recovered. Explicitly, \be
H_\alpha=\frac{1}{\kappa}{S_\alpha}^{\mu \nu} \eta_{\mu \nu},\ee
with ${S_\alpha}^{\mu \nu}=-{S_\alpha}^{\nu \mu}$ has the form
\cite{OP1} \be {S_\rho}^{\mu \nu} \stackrel {\rm def.}{=}
\frac{1}{4}\left({T_\rho}^{\mu
\nu}+{{T^\mu}_\rho}^\nu-{{T^\nu}_\rho}^\mu\right)-\frac{1}{2}\left({\delta_\rho}^\nu
{T_\theta}^{\mu \theta}-{\delta_\rho}^\mu {T_\theta}^{\nu
\theta}\right).\ee

Similarly, the explicit  form of the gravitational energy-momentum
\be E_\alpha=\frac{1}{2}\left[(e_\alpha \rfloor T^\beta) \wedge
H_\beta-T^\beta \wedge (e_\alpha \rfloor H_\beta) \right].\ee
Using (1), (17) and $T^\alpha={T_{\mu \nu}}^\alpha \vartheta^\mu
\wedge \vartheta^\nu=2 {\Gamma_{[\mu \nu]}}^\alpha \vartheta^\mu
\wedge \vartheta^\nu$ in (18) one can find \cite{LOP} \be
E_\alpha={t_\alpha}^\beta \eta_\beta, \qquad {t_\alpha}^\beta
=\frac{1}{2\kappa}\left(4{T_{\alpha \nu}}^\lambda
{S_\lambda}^{\beta \nu}-{T_{\mu \nu}}^\lambda {S_\lambda}^{\mu
\nu}{\delta_\alpha}^\beta\right).\ee By the same method one can
have \cite{LOP} \be \varepsilon_\alpha={j_\alpha}^\beta
\eta_\beta, \qquad  {j_\alpha}^\beta
=\frac{1}{2\kappa}\left(4{T_{\alpha \nu}}^\lambda
{S_\lambda}^{\beta \nu}-{T_{\mu \nu}}^\lambda {S_\lambda}^{\mu
\nu}{\delta_\alpha}^\beta+4{\Gamma_{\nu
\alpha}}^\lambda{S_\lambda}^{\beta \nu}\right).\ee

Now ${t_\alpha}^\beta$ is understood to be a true tensor since it
depends explicitly on the Weitzenb$\ddot{o}$ck connection
${\Gamma_{\nu \alpha}}^\lambda $ the current  ${j_\alpha}^\beta$
is a pseudotensor. Since the Weitzenb$\ddot{o}$ck  connection
${\Gamma_{\nu \alpha}}^\lambda $ represents the inertial effects
related to the choice of the frame, we see clearly that the origin
of the pseudotensor behavior of the usual energy-momentum
densities is that they include those inertial effects \cite{ALP}.

Taking into account the analogous expansion of the matter
energy-momentum, $\Sigma_\alpha = {\Sigma_\alpha}^\beta
\eta_\beta$, which introduces the energy-momentum tensor
${\Sigma_\alpha}^\beta $, and using (17) and (20), we easily
recover the field equation in tensor language (used, for example,
in [24]). Note that the conservation laws (15) and (16) coincide
when we put ${\Gamma_\alpha}^\beta=0$.  The last term in (21) then
disappears, whereas torsion reduces to the anholonomity 2-form,
$T^\alpha=F^\alpha= d\vartheta^\alpha$. We denote the
corresponding energy-momentum and superpotential with a tilde: \be
\widetilde{E}_\alpha=E_\alpha|_{{\Gamma_\alpha}^\beta=0}, \qquad
\widetilde{H}_\alpha=H_\alpha|_{{\Gamma_\alpha}^\beta=0}.\ee The
properties of these quantities and their use for the computation
of the total energy of the exact solutions was discussed in [22,
25]. Explicitly, one can have \cite{LOP} \be
\widetilde{H}_\alpha=\frac{1}{2\kappa}{\widetilde{\Gamma}}^{\beta
\gamma}\wedge  \eta_{\alpha \beta \gamma},\ee \be
\widetilde{E}_\alpha=\frac{1}{2}\left[(e_\alpha \rfloor
d\vartheta^\beta)\wedge \widetilde{H}_\beta-d\vartheta^\beta
\wedge (e_\alpha \rfloor \widetilde{H}_\beta)\right].\ee
\newsection{Total energy of Schwarzschild and Reissner Nordstr$\ddot{o}$m  solutions}
Lucas et al. \cite{LOP} have calculated the energy of three
different solutions  that reproduce the same metric which gives
the Schwarzschild metric. Also they have calculated the total
energy of Kerr metric. Here we are going to {\it generalized} this
calculation for another solutions that give the Schwarzschild and Reissner
Nordstr$\ddot{o}$m
metrics. We will calculate the total energy using the
tensorial expression of the energy-momentum.
\subsection{ Schwarzschild   metric}

Using the spherical local coordinates $(t,r,\theta, \phi)$,
Schwarzschild solution is described by the coframe components: \be
{{\vartheta^{{}^{{}^{\!\!\!\!\scriptstyle{S}}}}}{_{}{_{}{_{}}}}^{\alpha}}
=\left({\Lambda^\alpha}_\gamma\right)
\left({\Lambda'^\gamma}_\delta\right) \vartheta^{\delta}, \ee
where the coframe $\vartheta^{\delta}$ has the form \ba
\vartheta^{\hat{0}}\A=\A\frac{1}{\alpha}cdt,\quad\vartheta^{\hat{1}}=\alpha
dr, \quad \vartheta^{\hat{2}}=r d\theta, \quad
\vartheta^{\hat{3}}=r\sin\theta d\phi,\ \ where \nonu
\alpha \A=\A \left(1-\frac{2m}{r}\right)^{-\frac{1}{2}}, \qquad
and \qquad m=GM/c^2, \ea the matrices ${\Lambda^\alpha}_\gamma$
and ${\Lambda'^\gamma}_\delta $ are defined as \be
\left({\Lambda^\alpha}_\gamma\right)=\left( \matrix{ 1 &  0 & 0 &
0 \vspace{3mm} \cr  0  &  \sin\theta \cos\phi &  \cos\theta
\cos\phi & - \sin\phi \vspace{3mm} \cr 0  & \sin \theta \sin \phi&
\cos\theta \sin\phi & \cos\phi \vspace{3mm} \cr 0  & \cos\theta &
-\sin\theta  & 0 \cr }\right)\; ,\ee which is the global Lorentz
transformation and the local Lorentz transformation
$\left({\Lambda'^\alpha}_\gamma\right)$ has the form
\be
\left({\Lambda'^\gamma}_\delta\right) = \left( \matrix{ \beta &  \beta_1 \sin\theta
\cos\phi & \beta_1 \sin\theta \sin\phi &  \beta_1  \cos\theta \vspace{3mm} \cr
- \beta_1 \sin\theta \cos\phi & -1+\left(1-\beta \right)\sin^2\theta
\cos^2\phi &\left(1-\beta \right)\sin^2\theta \sin\phi \cos\phi
&\left(1-\beta \right)\sin\theta \cos\theta \cos\phi \vspace{3mm} \cr
- \beta_1 \sin\theta \sin\phi &\left(1-\beta \right) \sin^2\theta \sin\phi
\cos\phi &-1+\left(1-\beta \right)\sin^2\theta \sin^2\phi &\left(1-\beta \right)
\sin\theta \cos\theta \sin\phi \vspace{3mm} \cr - \beta_1
\cos\theta &\left(1-\beta \right)\sin\theta \cos\theta \cos\phi
&\left(1-\beta \right)\sin\theta \cos\theta \sin\phi  &-1+\left(1-\beta \right)\cos^2\theta \cr}\right), \ee
 where $\beta$ and $\beta_1$ have the form
 \be \beta = \frac{1}{\sqrt{1-\displaystyle\frac{2c_1}{r}}},
\qquad \qquad  \qquad \qquad \qquad
\beta_1=\sqrt{\displaystyle\frac{\frac{2c_1}{r}}{1-\displaystyle\frac{2c_1}{r}}}.
\ee where  $c_1$ is a constant. If we take tetrad (26), as well as
the trivial Weitzenb$\ddot{o}$ck connection
${\Gamma^\alpha}_\beta=0$ and substitute into (22) we finally get
\be \widetilde{H}_{\hat{0}} =\frac{r\sin \theta}{8\pi}
\left[\left(1-\sin^2\theta\cos^2\phi-\cos\theta\sin\phi\left[
1+\sin\theta\cos\phi\right]\right)\left\{ \beta-1\right\}
-\frac{2\beta}{\alpha} \right](d\theta\wedge d\phi).\ee If we
compute the total energy at a fixed time in the 3-space with a
spatial boundary 2-dimensional surface $\partial S =\{r = R,
\theta,\phi\}$ we obtain \be \widetilde{E} =\int_{\partial S}
\widetilde{H}_{\hat{0}}=\frac{R}{3} \left\{2+\beta
\left(1-\frac{3}{\alpha}\right)\right\}.\ee This case is similar
to the freely falling discussed in \cite{LOP,MFU} when $c_1\neq
0$, i.e., $E \neq M$ but here, the acceleration is not vanishing.
If we put the constant $c_1=0$ in Eq. (31),  then $\beta=1$ and
the energy will be the ADM, i.e., $E=M$ which is the case of the
proper tetrad \cite{LOP}. This is due to the fact that local
Lorentz transformation will be Minkowski metric, $o_{\alpha
\beta}=diag(+1,-1,-1,-1)$

Using the regularization framework which is based on the
covariance property, i.e., we will take into account the
Weitzenb$\ddot{o}$ck connection ${\Gamma^\alpha}_\beta \neq 0$ in
Eq. (17)  and calculate the necessary components we finally get
the superpotential \be H_{\hat{0}} =
\frac{r\beta\sin\theta\left(1-\displaystyle
\frac{1}{\alpha}\right)}{4\pi} (d\theta\wedge d\phi).\ee The total
energy of (31) thus has the form \be E=\int_{\partial S}
H_{\hat{0}}= R\beta\left(1-\displaystyle
\frac{1}{\alpha}\right)\cong
M\beta=M+O\left(\frac{1}{R}\right).\ee The non vanishing
components  needed to calculate the spatial momentum have the form
$\widetilde{H}_{\hat{\alpha}}=H_{\hat{\alpha}}, \
\hat{\alpha}=1,2,3$ have the form \ba
\widetilde{H}_{\hat{1}}=H_{\hat{1}} \A\cong \A
\frac{r\beta_1(1-\frac{1}{\alpha})\sin\theta\left[\sin\phi
\cos\phi\{1-\sin\theta\cos\theta\}-\cos^2\phi\sin^2\phi\right]}{4\pi}
(d\theta\wedge d\phi),\nonu
\widetilde{H}_{\hat{2}}=H_{\hat{2}} \A\cong \A
\frac{r\beta_1(1-\frac{1}{\alpha})\sin\theta\left[\cos\theta
\cos\phi\{\sin\theta\cos\phi-1\}-\cos\phi\sin\phi\sin^2\theta-\cos\theta\sin\theta\right]}{4\pi}
(d\theta\wedge d\phi), \nonu
\widetilde{H}_{\hat{3}}=H_{\hat{3}} \A\cong \A
\frac{r\beta_1(1-\frac{1}{\alpha})\sin^2\theta\left[\cos\theta
\cos\phi-\sin\theta\sin\phi\right]}{4\pi} (d\theta\wedge d\phi).
\ea Using Eqs. (34) we finally get the spatial momentum in the
form \be P_1=\int_{\partial S} H_{\hat{1}}=
R\beta_1\left(1-\displaystyle \frac{1}{\alpha}\right)\cong
M\beta_1=O\left(\frac{1}{R}\right),\qquad \qquad P_2=P_3=0.\ee
\subsection{ Reissner Nordstr$\ddot{o}$m metric }

Using the spherical local coordinates $(t,r,\theta, \phi)$,
Reissner Nordstr$\ddot{o}$m solution is described by the coframe
components: \be
{{\vartheta^{{}^{{}^{\!\!\!\!\scriptstyle{R}}}}}{_{}{_{}{_{}}}}^{\alpha}}
={\Lambda^\alpha}_\gamma {\Lambda''^\gamma}_\delta
\vartheta^{\delta}, \ee where the coframe $\vartheta^{\delta}$ has
the form\be
\vartheta^{\hat{0}}=\frac{1}{\alpha_1}cdt,\quad\vartheta^{\hat{1}}=\alpha_1
dr, \quad \vartheta^{\hat{2}}=r d\theta, \quad
\vartheta^{\hat{3}}=r\sin\theta d\phi,\ \ where \ \
\alpha_1=\left(1-\frac{2m}{r}+\frac{Q^2}{r^2}\right)^{-\frac{1}{2}},
\ee $\left({\Lambda^\alpha}_\gamma\right)$ is given by Eq. (27)
and $\left({\Lambda''^\gamma}_\delta\right) $ is defined as
\small{\be \left({\Lambda''^\gamma}_\delta\right) = \left(
\matrix{ \beta_2 &  \beta_3 \sin\theta \cos\phi & \beta_3
\sin\theta \sin\phi &  \beta_3  \cos\theta \vspace{3mm} \cr -
\beta_3 \sin\theta \cos\phi & -1+\left(1-\beta_2
\right)\sin^2\theta \cos^2\phi &\left(1-\beta_2
\right)\sin^2\theta \sin\phi \cos\phi &\left(1-\beta_2
\right)\sin\theta \cos\theta \cos\phi \vspace{3mm} \cr - \beta_3
\sin\theta \sin\phi &\left(1-\beta_2 \right) \sin^2\theta \sin\phi
\cos\phi &-1+\left(1-\beta_2 \right)\sin^2\theta \sin^2\phi
&\left(1-\beta_2 \right) \sin\theta \cos\theta \sin\phi
\vspace{3mm} \cr - \beta_3 \cos\theta &\left(1-\beta_2
\right)\sin\theta \cos\theta \cos\phi &\left(1-\beta_2
\right)\sin\theta \cos\theta \sin\phi  &-1+\left(1-\beta_2
\right)\cos^2\theta \cr}\right), \ee}
 where $\beta_2$ and $\beta_3$ have the form
 \be \beta_2 = \displaystyle\frac{1}{\sqrt{{1-\frac{2c_1}{r}+\frac{{c_2}^2}{r^2}
 }}},
\qquad \qquad  \qquad \qquad \qquad
\beta_3=\sqrt{\displaystyle\frac{{\frac{2
c_1}{r}+\frac{{c_2}^2}{r^2}}}{{1-\frac{2
c_1}{r}+\frac{{c_2}^2}{r^2}}}}, \ee with $c_2$ is another
constant. Take tetrad (36), as well as the trivial
Weitzenb$\ddot{o}$ck connection ${\Gamma^\alpha}_\beta=0$ and
substitute into (22) we finally get \be \widetilde{H}_{\hat{0}}
\frac{r\sin \theta}{8\pi}
\left[\left(1-\sin^2\theta\cos^2\phi-\cos\theta\sin\phi\left[
1+\sin\theta\cos\phi\right]\right)\left\{ \beta_2-1\right\}
-\frac{2\beta_2}{\alpha_1} \right] (d\theta\wedge d\phi).\ee In
general the acceleration of solution (36) is not vanishing. If we
compute the total energy at a fixed time in the 3-space with a
spatial boundary 2-dimensional surface $\partial S =\{r = R,
\theta,\phi\}$, we obtain \be   \widetilde{E} =\int_{\partial S}
\widetilde{H}_{\hat{0}}=\frac{R}{3} \left\{2+\beta_2
\left(1-\frac{3}{\alpha_1}\right)\right\}. \ee  When we put the
parameter $c_1=0$ and $c_2=0$ in Eq. (39), the energy will be
identical to  Reissner Nordstr$\ddot{o}$m \cite{Nar} because
$\beta_2=1$.

Using the regularization framework which is based on the covariance property, i.e.,
 we will take into account the  Weitzenb$\ddot{o}$ck connection ${\Gamma^\alpha}_\beta \neq
 0$ in Eq. (17)
 and calculate the necessary components we finally get the superpotential \be H_{\hat{0}} = \frac{r\beta_2\sin\theta}{4\pi}
\left(1-\displaystyle \frac{1}{\alpha_1}\right) (d\theta\wedge
d\phi).\ee The total energy of (42) thus has the form \be
E=\int_{\partial S} H_{\hat{0}}= R\beta_2 \left(1-\displaystyle
\frac{1}{\alpha_1}\right)=M-\frac{Q^2-2Mc_1}{2R}+O\left(\frac{1}{R^2}\right).\ee
 The  non vanishing components needed to calculate the spatial momentum have
of the form  $\widetilde{H}_{\hat{\alpha}}=H_{\hat{\alpha}}, \
\hat{\alpha}=1,2,3$ have the form \ba
\widetilde{H}_{\hat{1}}=H_{\hat{1}} \A\cong \A
\frac{r\beta_3(1-\frac{1}{\alpha_1})\sin\theta\left[\sin\phi
\cos\phi\{1-\sin\theta\cos\theta\}-\cos^2\phi\sin^2\phi\right]}{4\pi}
(d\theta\wedge d\phi),\nonu
\widetilde{H}_{\hat{2}}=H_{\hat{2}} \A\cong \A
\frac{r\beta_3(1-\frac{1}{\alpha_1})\sin\theta\left[\cos\theta
\cos\phi\{\sin\theta\cos\phi-1\}-\cos\phi\sin\phi\sin^2\theta-\cos\theta\sin\theta\right]}{4\pi}
(d\theta\wedge d\phi), \nonu
\widetilde{H}_{\hat{3}}=H_{\hat{3}} \A\cong \A
\frac{r\beta_3(1-\frac{1}{\alpha_1})\sin^2\theta\left[\cos\theta
\cos\phi-\sin\theta\sin\phi\right]}{4\pi} (d\theta\wedge d\phi).
\ea Using Eqs. (44) we finally get the spatial momentum in the
form \be P_1=\int_{\partial S} H_{\hat{1}}=
R\beta_3\left(1-\displaystyle \frac{1}{\alpha_1}\right)\cong
M\beta_3=O\left(\frac{1}{\sqrt{R}}\right),\qquad \qquad
P_2=P_3=0.\ee
\newsection{ Discussion and conclusion }

Teleparallel theory is considered as an essential part of
generalized non-Riemannian theories such as the Poincar$\acute{e}$
gauge theory \cite{Yi1} $\sim$ \cite{Kw} or metric-affine gravity
\cite{HMM}. Physics relevant to geometry may be related to the
teleparallel description of gravity \cite{HS,NH}. Within the
framework of metric-affine gravity, a stationary axially symmetric
exact solution of the vacuum field equations is obtained for a
specific gravitational Lagrangian by using {\it prolongation
techniques} (\cite{BH} and references therein).  Teleparallel
approach is used for positive-gravitational-energy proof
\cite{Me}. It is shown that {\it one} of the main  differences
between general relativity and teleparallel theory is that the
Weitzenb$\ddot{o}$ck connection which represents only inertial effects
related to the frame \cite{LOP}. Therefore, one can separate
gravitation from inertial effects. A tensorial expression for the
energy-momentum density of gravity is obtained in \cite{LOP} . A
covariant teleparallel approach naturally yields regularized
solutions for the energy and momentum due to the fact that the
frame-related inertial contribution to the conserved quantities is
always properly subtracted by the Weitzenb$\ddot{o}$ck connection.

In this study we show that in general, the total conserved
energy-momentum $\widetilde{P}_\alpha$ corresponding to
$\widetilde{H}_\alpha$  does not transform covariantly under a
change of frame. However, for local Lorentz transformations which
become global at spatial infinity, the total energy-momentum
transforms covariantly as a Lorentz vector. This is clear from the
two examples we have studied with two constants. These examples
reproduce Schwarzschild and Reissner Nordstr$\ddot{o}$m metrics.
The constants $c_1$ and $c_2$ of these solutions plays the role of
inertial as  shown in Eqs. (31) and (41) which makes the total
energy always unphysics. The only choice which makes the energy
always physics  is that $\beta=1$ for the first example, i.e.,
Schwarzschild solution and $\beta_2=1$ for the second example,
i.e., Reissner Nordstr$\ddot{o}$m solution. The choice $\beta=1$
leads to $c_1=0$ which makes the local Lorentz transformation (28)
has the form of Minkowski metric $o_{\alpha
\beta}=diag(+1,-1,-1,-1)$. When these condition is satisfied we
reproduce the case of proper tetrad discussed in \cite{LOP}. Same
discussion can be carried out for the choice $\beta_2=1$ but here
we also require $c_2=0$ .

 Therefore,
 we use the tensorial expression for the energy-momentum density of
gravity and calculate the total energy associated with the
Schwarzschild and Reissner Nordstr$\ddot{o}$m solutions. We show
by  calculations that the Weitzenb$\ddot{o}$ck connection acts as
a regularizing tool which separates the inertial energy-momentum
density, leaving the tensorial, physical energy-momentum density
of the system untouched. This is clear from Eq. (33) in which the
energy is given by $E \cong M \beta\cong M(1+\frac{c_1}{R}) \cong
M$. This shows that the terms that will contributes to the energy
is of order $O(1/R)$. In this case, i.e., Schwarzschild,  we do
not need terms of $O(1/R)$. On the other hand, for the  Reissner
Nordstr$\ddot{o}$m solution the energy is given by Eq. (43), i.e,
$E\cong \beta_2(M-\frac{Q^2}{2R})\cong
(1+\frac{c_1}{R}-\frac{{c_2}^2}{2R^2})(M-\frac{Q^2}{2R})\cong
M-\frac{Q^2-2Mc_1}{2R}+O(1/R^2)$. In this case, the constant $c_1$
contributes the total energy. When this constant, i.e., $c_1=M/2$
the value of energy will be consistent \cite{NS}. Finally we show
that the components needed to calculate  the spatial momentum have
the same form either we put the  Weitzenb$\ddot{o}$ck connection
trivial or non trivial. As is clear from Eqs (35) and (45) the
components of spatial momentum associated with  Schwarzschild and
Reissner Nordstr$\ddot{o}$m solutions are in agreement with the
previous results \cite{ORP,NS}. Finally we show that the inertial
has no effect on the calculation of the spatial components as
explained in the both examples studied in paper.

\end{document}